\documentclass[aps,prl,twocolumn,groupedaddress,showpacs]{revtex4-1}

\usepackage{dcolumn}
\usepackage{graphicx,amssymb,amsmath}
\usepackage{bm}
\usepackage{natbib}
\usepackage{epstopdf}
\begin{document}

\title{ Non-perturbative Solution of the Unitary, $N$-Orbital Anderson Model}
\author{Shimul Akhanjee$\dagger$ }
\email[]{sakhanjee@bnl.gov}
\affiliation{Condensed Matter Theory Laboratory, RIKEN, Wako, Saitama, 351-0198, Japan}

\date{\today}

\begin{abstract}The mobility edge is extracted from a non-perturbative analysis of F. Wegner's real matrix ensemble (RME), $N$-orbital model of electrons with broken time-reversal invariance moving in random potential. The replicon fluctuations around the zero-dimensional (0D), single metallic grain saddlepoint, are shown to destabilize diffusive regime beyond the low energy limit yielding a precise criteria of localization in terms of a single scaling parameter. 
\end{abstract}


\maketitle

\emph{Introduction}-- The mobility edge problem for disordered electron systems has an extensive history dating back to the seminal work of Anderson in 1958, who designated a criteria of localization for the insulating phase based on the existence of discrete levels in the energy spectrum that are accompanied by normalizable eigenfunctions\cite{anderson}. Localization signatures are found within the squared, impurity averaged Green's function, $\mathop {\lim }\limits_{t \to \infty } {\left| {G(r,r';t)} \right|^2}$\cite{anderson,*econ1970,*gangfour}, or more generally, higher moments of the wave function which also contain distinct non-analyticities\cite{lifshitsdisorderbook}. Although there are a variety of perturbative methods for disordered systems, all seem to breakdown near the transition point (aside from the Bethe lattice), and the exact mobility edge trajectory has not yet been determined.

An effective field theoretic approach was introduced by F. Wegner and collaborators\cite{wegner1979}, that built on earlier ideas\cite{nitzanPRB1977,*aharonyimry1977} developed from a replica field theory. Wegner identified a continuous symmetry of the $N$-orbital version of Anderson's tight-binding model and a formulated the mobility edge problem within the same mathematical language of traditional symmetry breaking transitions. Moreover, Wegner introduced a replicated generating functional $ F$, that can in principle, compute the exact correlators of the disordered electron gas, $G(r,r';t)$ and $K(r_1,r'_1,z_1;r_2,r'_2,z_2)$, provided that the non-perturbative aspects of the disorder average in the replica limit $n\to0$ can be taken correctly\cite{wegnerschafer1980}. Physically, the $N$-orbital model can be understood as a system of metallic grains coupled to its nearest neighbors, forming a $D$-dimensional lattice. The large $N$ limit is equivalent to the random phase approximation, or a mean-field theory of the Anderson model. 

An exact description of the mobility edge has been elusive, in part, due to the prevailing need to address the replica symmetry-breaking substructure of saddlepoint manifold which occurs at higher energies away from the diffusive regime. Although Wegner's perturbative approximation cannot access the transition point, it can however confirm particular conductance scaling forms. In this paper, an evaluation of $F$ is obtained by exploiting recent non-perturbative methods developed for spin-glass replica field theories with unitary symmetry or broken time-reversal invariance\cite{spherePRL}. Previous 0D studies have focused on level correlations within the ergodic regime $\omega<<1$\cite{efetovbook,*kanzieperPRL2002,*lerner} (in units of the mean level spacing), however the treatment here shall attempt to observe signatures of localized states arising from destabilizing replicon spatial fluctuations at higher energies.

\emph{Model}-- Consider the one-particle, single band, tight-binding model of Wegner\cite{oppermanwegner1979}, with $N$ orbitals per site, ${\left| r,\alpha \right\rangle }$, $\alpha =1,\dots,N$ on a regular lattice. The pair of indices $(r,\alpha)$ will be abbreviated by $x$, 
\begin{equation}
H = \sum\limits_{xx'} {\frac{1}{{\sqrt N }}{f_{xx'}}} \left| x \right\rangle \left\langle {x'} \right|
\label{eq:ohamilt}
\end{equation}
and defined within a basis that is invariant under the local $\mathbb{Z}^2$ gauge transformation,
\begin{equation}
\left| {r,\alpha } \right\rangle  \to {\sigma _{r,\alpha }}\left| {r,\alpha } \right\rangle \,\,\,\,\,\,\,\,\,\,\,\,\,\,\,{\sigma _{r,\alpha }} =  \pm 1\,\,\,\,\,\,\,\,\,\,
\label{eq:rmegauge}
\end{equation}
The matrix elements $f_{xx'}$ form a real symmetric matrix or RME ensemble, distributed according to a normalized Gaussian weight $P[f_{xx' }]$, with ensemble averages taken over independent matrix elements: $\left[\kern-0.15em\left[ A \right]\kern-0.15em\right] = \int {D[} f]P[f]A[f]$, assuming $\left[\kern-0.15em\left[ {{f_{xx' }}} 
 \right]\kern-0.15em\right] = 0$ and $\left[\kern-0.15em\left[ {{f_{{x_1}{x_1}'}}{f_{{x_2}{x_2}'}}} 
 \right]\kern-0.15em\right] = \left( {{\delta _{{x_1}{x_2}'}}{\delta _{{x_2}{x_1}'}} + \kappa{\delta _{{x_1}{x_2}}}{\delta _{{x_1}'{x_2}'}}} \right){M_{{r_1-r_1}'}}$, with disorder width ${E_0^2} = 4\sum\nolimits_{r'} {{M_{r-r'}}} $ and $\kappa=1$. It is necessary to define the normalized inverse, ${\left( {{M^{ - 1}}} \right)_{r-r'}} = 4{w_{r-r'}}/E_0^2$, such that $\sum\nolimits_{r'} {{w_{r - r'}}}  = 1$.

Wegner has also introduced the phase invariant ensemble (PIE), with $\kappa =0$ and eigenstates having $U(1)$ gauge symmetry\cite{oppermanwegner1979}, 
\begin{equation}
\left| {r,\alpha } \right\rangle  \to e^{i\phi_{r,\alpha }}\left| {r,\alpha } \right\rangle \,\,\,\,\,\,\,\,\,\,\,\,\,\,\,\phi_{r,\alpha }\,\, {\in \mathbb{R}}
\label{eq:piegauge}
\end{equation}

The averaged correlators of the RME (PIE) ensemble can be expressed as the expectation values of products of real (complex) valued spin vectors $\vec S(x)$, with components that are generated by the replica trick\cite{edanderson}, $S_a$, $a,\dots,n$, taken in the limit $n\to0$\cite{wegnerschafer1980}:
\begin{equation}
 G(r,r';z_\pm)=\left[\kern-0.15em\left[ {\left\langle r \right|{{({z_\pm} - H)}^{ - 1}}\left| {r'} \right\rangle } 
 \right]\kern-0.15em\right] =  \pm i\left\langle {{S_a}(r){S_a}(r')} \right\rangle 
 \end{equation}
\begin{equation}
\begin{array}{l}
 K(r_1,r'_1,z_+;r_2,r'_2,z_-)= \\
\left[\kern-0.15em\left[ {\left\langle {{r_1}} \right|{{({z_+} - H)}^{ - 1}}\left| {{r_1}'} \right\rangle \left\langle {{r_2}} \right|{{({z_-} - H)}^{ - 1}}\left| {{r_2}'} \right\rangle } 
 \right]\kern-0.15em\right] \\ 
  =  \left\langle {{S_a}({r_1}){S_a}({r_1}'){S_b}({r_2}){S_b}({r_2}')} \right\rangle  \\ 
 \end{array}
 \label{eq:gdef}
\end{equation}
with energy and frequency, $z_{\pm}\equiv E(k)\pm (\omega/2 + i\delta)$, for $E(k),\omega,\delta{\in \mathbb{R}}$. In terms of standard transport quantities, the mean-free path is given by $l=v_F\tau_{el}$, where $v_F$is the Fermi velocity and $\tau_{el}$ is the momentum relaxation rate within the Drude picture of electronic scattering, allowing the disorder strength to be expressed as $\eta \equiv E_0^2\pi \mathcal{N}(\epsilon_F) \propto 1/\tau_{el}$, with the density of states at the Fermi level $\mathcal{N}(\epsilon_F)$.
Next, in order to the define a suitable functional integral, the random tight-binding system can be reformulated in terms of a continuous, spin-glass Hamitonian,
\begin{equation}
\mathcal{H}_{sg} = \mp i \sum\nolimits_{r>r'}S_a (r) f_{r,r'} S_a (r') + z_\pm \sum\nolimits_r S_a^2 (r)
\label{eq:s1}
\end{equation}
allowing the averages defined through Eq.(\ref{eq:gdef}) to be expressed as,
\begin{equation}
\left\langle A \right\rangle  = \frac{{\int {D[\vec S]]A[\vec S]\exp \left( { - {\mathcal{H}_{sg}}} \right)} }}{{\int {D[\vec S]\exp \left( { - {\mathcal{H}_{sg}}} \right)} }}
\end{equation}
\emph{Wegner Functional}-- After integrating out $f_{r,r'}$, a four spin interaction is decoupled by a Hubbard-Stratonovich transformation and the $S_a$'s are integrated out to obtain the celebrated Wegner functional of interacting matrices;
\begin{equation}
\begin{aligned}
&\mathcal{Z}_n = \int {dQ\exp \left( {{F}(Q)} \right)}\\
\label{eq:Qform}
 &F(Q) = - \frac{N}{2} \times  \\ 
 &\sum\limits_{rr'} {{w_{r-r'}}{\rm Tr}\left( {Q(r)Q(r')} \right)}  + \sum\limits_r {{\rm Tr}\ln \left( {\hat {\rm I}\frac{z_\pm}{E_0} - Q(r)} \right)}  
\end{aligned}
\end{equation}
where $\hat {\rm I}$ is the identity matrix and the trace operation is taken in replica space. When the spins of Eq.(\ref{eq:s1}) interact with an infinite-ranged potential, the resulting matrix interactions are short-ranged, yielding a 0D replica field theory which was exactly solved in reference\cite{spherePRL} in the context of the spherical spin glass problem. However, for the $N$-orbital model studied here the spherical constraint will be supplanted by a saddle point solution for $Q$ at each lattice site, or in the continuum limit, an integration path in $Q$-space that must be deformed such that it passes through the saddlepoint, which depends on the choice of $z_\pm $. The relevant parts of the solution in ref. \cite{spherePRL} will be reviewed in the next section and the fluctuations about the 0D saddlepoint will be studied thereafter.

\emph{Exact 0D solution}-- The 0D limit, is a single metallic grain, where the primary complication of concern is the non-perturbative evaluation of ${F}$ in the $n\to 0$ limit. Therefore, the key technical advancement applied here is described in the reference\cite{spherePRL}, where a Jacobian of $ Q$ in terms of its eigenvalues $\lambda_i$, called the Vandermonde determinant: $d Q = \prod\nolimits_{1 \le j \le k \le n} {({\lambda _k} - {\lambda _j})}$
\cite{mehta}, allows Eq.(\ref{eq:Qform}) to be transformed into a Coulomb gas partition function taken in the limit $n\to 0$:  
\begin{equation}
\mathcal{Z}_n = \int \exp \left[  \mathcal{H}_{cg} (\lambda)\right] d \lambda_1 \ldots d \lambda_n
\end{equation}
containing a new Hamiltonian $\mathcal{H}_{cg}$ describing a one-dimensional gas of particles with logarithmic interactions, often encountered in random matrix theory\cite{mehta},
\begin{equation}
\mathcal{H}_{cg} (\lambda) = \sum\nolimits_i {V (\lambda_i)}  + \beta\sum\nolimits_{i > j} {\ln \left| {\lambda _i  - \lambda _j } \right|} 
\label{eq:cgham}
\end{equation}
coupling each charge to the single particle potential,
\begin{equation}
V (\lambda) = \frac{N}{2} \ln \left[ z - \lambda \right] + \frac{N}{2E_0^2 } \lambda^2
\label{eq:s8}
\end{equation}
Fixing $\beta=2$ corresponds to the unitary ensemble\cite{mehta,*FandW,*okamoto1981}; the partition function for this system was shown to be equivalent to the $\tau^{IV}$ function of Painlev\'e systems\cite{spherePRL}, 
\begin{equation}
\begin{aligned}
\tau^{IV}[n]&= \frac{1}{C} \int_{-\infty}^t dx_1 \cdots \int_{-\infty}^t dx_n \times\\
&\prod_{j=1}^n e^{-x_j^2} (t-x_j)^{\mu} \prod_{1 \leq j <k \leq n}  (x_k-x_j)^2
\end{aligned}
\label{eq:FWint1}
\end{equation}

The connection to the 0D Wegner functional can be obtained by taking $x_j \rightarrow y_j \sqrt{N/2E_0^2  }$, and $t \rightarrow z \sqrt{N/2E_0^2  }$, yielding 
\begin{equation}
\begin{aligned}
\mathcal{Z}_n&= (N/2E_0^2 )^{ (n (n-1)/2+n+\mu)/2} \int_{-\infty}^t dy_1 \cdots \int_{-\infty}^t dy_n\times \\
&\prod_{j=1}^{n}e^{-\frac{N}{2E_0^2 }y_j^2}  (z -y_j)^{\mu} \prod_{1 \leq j<k \leq n}(y_k-y_j)^2
\label{eq:integ1}
\end{aligned}
\end{equation}
for which the recurrence relations, for large $\mu=-N/2$ were explicitly solved to yield the following \cite{spherePRL},
\begin{equation}
\begin{aligned}
 {\tau ^{IV}}&[n] = {2^{ - n(n - 1)}}{\pi ^{ - n/2}} \\ 
 &{\left( {\frac{{\pi \left( {t - \sqrt {{t^2} - 8N} } \right)}}{{\sqrt {{t^2} - 8N} }}} \right)^{n/2}}\exp \left[ {\frac{{nN}}{8}\left( {t\sqrt {{t^2} - 8N} } \right)} \right] \times  \\ 
 &\exp \left[ { 8N\ln \left[ {\frac{1}{4}\left( {t - \sqrt {{t^2} - 8N} } \right)} \right] - \frac{N}{2} - 9{t^2}} \right] \times  \\ 
 &{\left( {1 + \frac{N}{4}\left( {\frac{t}{{\sqrt {{t^2} - 8N} }} - 1} \right)} \right)^{n(n - 1)/2}}  
 \end{aligned}
\label{eq:tau4}
\end{equation}

After completing the replica calculation by taking the derivative, 
\begin{equation}
 {F}= - \left[\kern-0.15em\left[ {\ln Z} \right]\kern-0.15em\right]
=\mathop {\lim }\limits_{n \to 0} \left ( {{\mathcal{Z}_n - 1}} \right)/n = {\left. {d{\mathcal{Z}_n}/dn} \right|_{n = 0}}.
\label{eq:replica}
\end{equation}
the dominant contributions to the functional become,
\begin{equation}
\begin{aligned}
&{F}(x) = N \times \\
&\left[ { -\frac{1}{2} + \frac{x^2}{8} - \frac{x}{8}\sqrt {{x^2} - 8}  - \ln \left( {\frac{{\sqrt N }}{4}\left( {x - \sqrt {{x^2} - 8} } \right)} \right)} \right] 
\end{aligned}
\label{eq:exfunc}
\end{equation}
where $x=z_{\pm}/E_0$. Upon comparison with Eq.(\ref{eq:Qform}), the 0D saddlepoint $Q$'s in the $n \to 0$ limit are taken to be
\begin{equation}
{Q_0^\pm} = (z_{\pm}/E_0 - \sqrt {{z_{\pm}^2/E_0^2} - 8}) 
\end{equation}
Accordingly, this agrees with the predicted saddlepoint structure for $r$-independent $Q$'s discussed by Schafer and Wegner\cite{wegnerschafer1980}, such that at $\omega\to 0 $, $Q_0$ is bounded by $ E/E_0 =\pm \sqrt{8}$. 

\emph{Spatial fluctuations about 0D}-- The low energy behavior of the Goldstone or replicon propagator, $K(q,\omega) \sim 1/(-i\omega +Dq^2)$ should be consistent with the well known scaling theory of localization\cite{gangfour}. However, as we will demonstrate here, in three-dimensions (3D), at small enough $E/E_0$, the onset of a negative gap should destabilize the ergodic regime into strong localization, where the dc conductivity should vanish. A precise criterion where this occurs can be determined by examining dynamical fluctuations about the exact 0D saddlepoint, in the $n\to 0$ limit. Unlike earlier approaches, our functional is valid at higher energies since the log term of $F(Q)$ is treated non-perturbatively, yielding a solution very different from the typical Wigner-Dyson correlations at lower energies\cite{efetovbook,kravtsov}. First, consider a fluctuating matrix field $Q'(\vec r)=Q(\vec r)-Q_0$ with coordinates shifted to the center of mass frame, $\vec r= \vec R+\vec\rho/2$, $\vec r'=\vec R-\vec \rho/2$ and $\vec r-\vec r'=\vec \rho$. In order to examine the dynamics, the matrix field fluctuations can be expanded out to the 2nd order in gradients as follows,
\begin{equation}
\begin{array}{l}
 Q'(\vec r) = Q'(\vec R) + \frac{\vec \rho }{2}\vec \nabla Q'(\vec R) + \frac{1}{2}{\left( {\frac{\vec \rho }{2}} \right)^2}{\vec \nabla ^2}Q'(\vec R) \\ 
 Q'(\vec r') = Q'(\vec R) - \frac {\vec \rho }{2}\vec \nabla Q'(\vec R) + \frac{1}{2}{\left( {\frac{\vec \rho }{2}} \right)^2}{\vec\nabla ^2}Q'(\vec R) \\ 
 \end{array}
\end{equation}
The primary contributions to the replicon modes should only depend on the non-vanishing 2nd derivative with respect to fluctuations, 
\begin{equation}
\begin{aligned}
\delta F(Q) &= F(Q(\vec r)) - F({Q_0}) \\
&=\sum\limits_{\rho ,\vec R} {Q'(\vec R){{\left[ {\frac{{\delta F}}{{\delta Q'(\vec R)\delta Q'(\vec R)}}} \right]}_{Q' = {Q_0}}}} Q'(\vec R)
\end{aligned}
\end{equation}
After taking the appropriate functional derivatives, the first term of Eq.(\ref{eq:Qform}) yields,
\begin{equation}
\delta F_1(Q) = -\sum\limits_{\rho ,\vec R} {Q'(\vec R)\left[ {2{w_\rho } + 4{w_\rho }{{\left( {\frac{\rho }{2}} \right)}^2}{\vec \nabla ^2}} \right]} Q'(\vec R)
\end{equation}
and by defining the mobility constant, $1/\mu_\rho = \left\langle {{\rho ^2}} \right\rangle  = \sum\limits_\rho  {{w_\rho }{\rho ^2}}$, the sum over $\rho$ yields:
\begin{equation}
\delta F_1(Q) = -\sum\limits_{\vec R} {Q'(\vec R)\left[ {2 + \mu_\rho^{-1}{\vec\nabla ^2}} \right]} Q'(\vec R)
\end{equation}
Next, we can perform a similar expansion of the log term of Eq.(\ref{eq:Qform}),
\begin{equation}
\delta {F_2}(Q) =   \sum\limits_{\vec R} {Q'(\vec R)\left[ {{{\left( {E/{E_0} - {Q_0}} \right)}^{ - 2}}} \right]} Q'(\vec R)
\end{equation}
and combining both expressions, the integrated replicon propagator becomes:
\begin{equation}
K(z_\pm) = \int_0^{{k _F}} {\frac{{{d^d}q}}{{\mu_\rho^{-1}{q^2} -2+ {{(z_\pm/{E_0} - {Q_0})}^{ - 2}}}}}
\label{eq:Kint} 
\end{equation}
It should be noted that $q$ is not the single particle momentum from the dispersion $E(k)=\hbar^2 k^2/2m-\epsilon_F$, rather it is Fourier component of the center of mass coordinate, $\vec R$, of which must be summed over the lattice or in the continuum limit, integrated over space. The full conductivity $\sigma$ should be obtained from summing over both advanced and retarded poles, $z_\pm$, however since $K(E)$ is sufficiently long ranged, the short-ranged contributions arising from the $s$ summation can be neglected\cite{wegnerPRB1979}. Taking the Fourier transform of integrand in Eq.(\ref{eq:Kint}),
\begin{equation}
S(R) = \frac{1}{{4{\pi ^2}}}\sum\limits_{s =  \pm } {\int {\frac{{{e^{ - iqR}}}}{{{\mu ^{ - 1}}{q^2} - 2 + {{({z_s}/{E_0} - {Q_0})}^{ - 2}}}}{d^d}q} } 
\end{equation}
the conductivity becomes, 
\begin{equation}
\sigma (\omega ,E) \propto {\omega ^2}\int {{R ^2}S(R ){d^d}R } 
\end{equation}

Thus, in order to ascertain the bounds on $\sigma$, as in case of weak-localization, the divergence of the integral (\ref{eq:Kint}) indicates the onset of localization at critical value of the scaling parameter $t_c=E_c/E_0$, in the dc limit $\omega\to 0$, which is independent of the sum over both saddlepoints, $z_\pm$.

\emph{The mobility edge}-- In 3D we have,
\begin{equation}
\frac{K(E)}{4\pi} = \mu_\rho k_F- \left( {\mu_\rho^{3/2}\Delta } \right)\left\{ \begin{array}{l}
 {\tan ^{ - 1}}( {k_F}/\Delta \sqrt {\mu_\rho})\,\,\,\,\,\,\,{\Delta ^2} > 0 \\ 
 {\tanh ^{ - 1}}({k_F}/\Delta \sqrt {\mu_\rho})\,\,\,\,\,{\Delta ^2} < 0 \\ 
 \end{array} \right.
\end{equation}
where the replicon energy gap is defined as,
\begin{equation}
{\Delta ^2} = -(2 - {({(E/{E_0})^2} - 8)^{ - 1}})
\label{eq:mobedge}
\end{equation}
for which ${\Delta ^2}<0$, within $(E/{E_0})\in (-\sqrt{8},\sqrt{8})$. Therefore, $K(E)$ clearly suffers from a strong negative divergence in the regime ${k_F}/\sqrt {\mu_\rho}\Delta \ge 1$, which generates the condition for localization,
\begin{equation}
\frac{{k_F^2}}{{\mu_\rho(2 - {{({{(E/{E_0})}^2} - 8)}^{ - 1}})}} \ge 1
\end{equation}
or the mobility edge trajectory in terms of a single scaling parameter $E/ E_0$. There are several quantitative limits of interest for which the consistency of condition (\ref{eq:mobedge}) can be confirmed. At weaker disorder strength (larger $E/ E_0$), $K(E)>0$ is simply offset by small, negative corrections. In the strong disorder limit $E_0 \to \infty$, $K(E)$ acquires a large negative gap in a manner similar to the de Almeida-Thouless instability of mean-field spin glass systems\cite{dealmeida}. An important distinction here is that I considered spatial fluctuations about the 0D saddlepoint rather than replica symmetry-breaking fluctuations about the replica symmetric saddlepoint. 

In order to make a connection with known results of the single orbital Anderson model with unitary symmetry, the scaled critical width was determined numerically as $E_0^c \approx 18.3$\cite{slevinPRL1997}. In the $N$-orbital model, the mobility inverse $1/\mu_\rho$ can be re-expressed in terms of transport parameters by dimensional analysis: $1/\mu_\rho \propto A E_0^2 $, for some positive constant A, and $E_0^c$, follows from $E \to 0$ in Eq.(\ref{eq:mobedge}), yielding $(E_0^c)^2 = 15/(8A)$.

\emph{Discussion}--  My primary focus here was to demonstrate critical localization within the framework of a 0D replica field theory beyond the low energy limit. Consequently, the precise determination of the transition point was not possible with earlier low energy, non-linear sigma model approaches that are only valid within the diffusive regime\cite{kravtsov}. Hence, I have presented the gradient expansion of spatial fluctuations about the single grain saddlepoint as a new mean-field, hydrodynamic approach which can access the breakdown of the ergodic regime into localized states within a mobility edge trajectory. These spatial correlations can be quantitatively bounded within a Ginzburg criterion, where the variance of the spatial fluctuations should be compared to the square of its average value. The negative divergence of the integrated replicon propagator clearly demonstrates a breakdown of the dc conductivity and the fundamental necessity of restoring non-mean-field corrections, which are beyond the scope of the gradient expansion. 

The single grain saddlepoint utilized here is a mathematically exact starting point for studying Anderson localization, facilitating investigations of complexities severely lacking in conceptual understanding such as the replica symmetry-breaking substructure of the localized regime\cite{parisiPRL}, which is common to both spin-glass and localization phenomena. Hence, the methods introduced here can be further refined to include fluctuations in both real and replica spaces by taking the expression of the 0D saddlepoint solution $\tau^{IV}[n]$ given by Eq.(\ref{eq:tau4}), which precedes the $n\to0$ limit, and including replica symmetry breaking dynamics, thereby developing a stability criterion along the lines of de Almeida and Thouless\cite{dealmeida}. Thus, for a given symmetry class of disordered systems, a different 0D $dQ$ Jacobian will lead to a different type of Coulomb gas ensemble, which can in principle can be evaluated in terms of random matrix averages. 

Recently, Anderson localization has played a central role in determining the robustness of surface states of topological insulators\cite{hasankane}. It particular, it has been difficult to reconcile the topological terms that represent magnetic field or Berry curvature effects\cite{pruiskenNUCB1984} within the sigma model approach and access higher energy localized states. Beyond the approximation of a spatially uniform saddlepoint, the model studied here at $\beta=2$ is of the unitary symmetry class, which is the simplest case to consider because it avoids the complexities of Pfaffians that arise in the orthogonal ensemble $\beta=1$. Thus, an extension of the methods here may offer an alternative route towards a solution of the the elusive the integer quantum Hall plateau transition, which is a mobility edge problem in the presence of a magnetic field through a modification of ${F}$ to include a topological term. In that case the 2D quantum hall system is delocalized at a critical value of the topological winding number, $\Theta = \pi$ in the diffusive regime, however a more complete description would involve localization criterion that includes both $\Theta$ and the scaling parameter $E/E_0$ into an inequality similar to Eq.(\ref{eq:mobedge}). 

I am grateful to Prof. J. Rudnick for our earlier collaboration that inspired this work and for helpful discussions with Dr. A. Furusaki. I thank Dr. A. M. Tsvelik for critical comments and important references. I acknowledge support from the RIKEN FPR program.

$\dagger$ Present Address: Condensed Matter Theory Group, CMPMSD, Brookhaven National Laboratory, Upton, NY 11973


\begin{thebibliography}{24}%
\makeatletter
\providecommand \@ifxundefined [1]{%
 \@ifx{#1\undefined}
}%
\providecommand \@ifnum [1]{%
 \ifnum #1\expandafter \@firstoftwo
 \else \expandafter \@secondoftwo
 \fi
}%
\providecommand \@ifx [1]{%
 \ifx #1\expandafter \@firstoftwo
 \else \expandafter \@secondoftwo
 \fi
}%
\providecommand \natexlab [1]{#1}%
\providecommand \enquote  [1]{``#1''}%
\providecommand \bibnamefont  [1]{#1}%
\providecommand \bibfnamefont [1]{#1}%
\providecommand \citenamefont [1]{#1}%
\providecommand \href@noop [0]{\@secondoftwo}%
\providecommand \href [0]{\begingroup \@sanitize@url \@href}%
\providecommand \@href[1]{\@@startlink{#1}\@@href}%
\providecommand \@@href[1]{\endgroup#1\@@endlink}%
\providecommand \@sanitize@url [0]{\catcode `\\12\catcode `\$12\catcode
  `\&12\catcode `\#12\catcode `\^12\catcode `\_12\catcode `\%12\relax}%
\providecommand \@@startlink[1]{}%
\providecommand \@@endlink[0]{}%
\providecommand \url  [0]{\begingroup\@sanitize@url \@url }%
\providecommand \@url [1]{\endgroup\@href {#1}{\urlprefix }}%
\providecommand \urlprefix  [0]{URL }%
\providecommand \Eprint [0]{\href }%
\providecommand \doibase [0]{http://dx.doi.org/}%
\providecommand \selectlanguage [0]{\@gobble}%
\providecommand \bibinfo  [0]{\@secondoftwo}%
\providecommand \bibfield  [0]{\@secondoftwo}%
\providecommand \translation [1]{[#1]}%
\providecommand \BibitemOpen [0]{}%
\providecommand \bibitemStop [0]{}%
\providecommand \bibitemNoStop [0]{.\EOS\space}%
\providecommand \EOS [0]{\spacefactor3000\relax}%
\providecommand \BibitemShut  [1]{\csname bibitem#1\endcsname}%
\let\auto@bib@innerbib\@empty
\bibitem [{\citenamefont {Anderson}(1958)}]{anderson}%
  \BibitemOpen
  \bibfield  {author} {\bibinfo {author} {\bibfnamefont {P.~W.}\ \bibnamefont
  {Anderson}},\ }\href@noop {} {\bibfield  {journal} {\bibinfo  {journal}
  {Phys. Rev.}\ }\textbf {\bibinfo {volume} {109}},\ \bibinfo {pages} {1492}
  (\bibinfo {year} {1958})}\BibitemShut {NoStop}%
\bibitem [{\citenamefont {Economou}\ and\ \citenamefont
  {Cohen}(1970)}]{econ1970}%
  \BibitemOpen
  \bibfield  {author} {\bibinfo {author} {\bibfnamefont {E.~N.}\ \bibnamefont
  {Economou}}\ and\ \bibinfo {author} {\bibfnamefont {M.~H.}\ \bibnamefont
  {Cohen}},\ }\href@noop {} {\bibfield  {journal} {\bibinfo  {journal} {Phys.
  Rev. Lett.}\ }\textbf {\bibinfo {volume} {25}},\ \bibinfo {pages} {1445}
  (\bibinfo {year} {1970})}\BibitemShut {NoStop}%
\bibitem [{\citenamefont {Abrahams}\ \emph {et~al.}(1979)\citenamefont
  {Abrahams}, \citenamefont {Anderson}, \citenamefont {Licciardello},\ and\
  \citenamefont {Ramakrishnan}}]{gangfour}%
  \BibitemOpen
  \bibfield  {author} {\bibinfo {author} {\bibfnamefont {E.}~\bibnamefont
  {Abrahams}}, \bibinfo {author} {\bibfnamefont {P.~W.}\ \bibnamefont
  {Anderson}}, \bibinfo {author} {\bibfnamefont {D.~C.}\ \bibnamefont
  {Licciardello}}, \ and\ \bibinfo {author} {\bibfnamefont {T.~V.}\
  \bibnamefont {Ramakrishnan}},\ }\href@noop {} {\bibfield  {journal} {\bibinfo
   {journal} {Phys. Rev. Lett.}\ }\textbf {\bibinfo {volume} {42}},\ \bibinfo
  {pages} {673} (\bibinfo {year} {1979})}\BibitemShut {NoStop}%
\bibitem [{\citenamefont {Lifshits}\ \emph {et~al.}(1988)\citenamefont
  {Lifshits}, \citenamefont {Gredeskul},\ and\ \citenamefont
  {Pastur}}]{lifshitsdisorderbook}%
  \BibitemOpen
  \bibfield  {author} {\bibinfo {author} {\bibfnamefont {I.~M.}\ \bibnamefont
  {Lifshits}}, \bibinfo {author} {\bibfnamefont {S.~A.}\ \bibnamefont
  {Gredeskul}}, \ and\ \bibinfo {author} {\bibfnamefont {L.~A.}\ \bibnamefont
  {Pastur}},\ }\href@noop {} {\emph {\bibinfo {title} {Introduction to the
  Theory of Disordered Systems}}}\ (\bibinfo  {publisher} {John Wiley and
  Sons},\ \bibinfo {address} {New York},\ \bibinfo {year} {1988})\BibitemShut
  {NoStop}%
\bibitem [{\citenamefont {Wegner}(1979{\natexlab{a}})}]{wegner1979}%
  \BibitemOpen
  \bibfield  {author} {\bibinfo {author} {\bibfnamefont {F.}~\bibnamefont
  {Wegner}},\ }\href@noop {} {\bibfield  {journal} {\bibinfo  {journal} {Z.
  Phys. B}\ }\textbf {\bibinfo {volume} {35}},\ \bibinfo {pages} {207}
  (\bibinfo {year} {1979}{\natexlab{a}})}\BibitemShut {NoStop}%
\bibitem [{\citenamefont {Nitzan}\ \emph {et~al.}(1977)\citenamefont {Nitzan},
  \citenamefont {Freed},\ and\ \citenamefont {Cohen}}]{nitzanPRB1977}%
  \BibitemOpen
  \bibfield  {author} {\bibinfo {author} {\bibfnamefont {A.}~\bibnamefont
  {Nitzan}}, \bibinfo {author} {\bibfnamefont {K.~F.}\ \bibnamefont {Freed}}, \
  and\ \bibinfo {author} {\bibfnamefont {M.~H.}\ \bibnamefont {Cohen}},\
  }\href@noop {} {\bibfield  {journal} {\bibinfo  {journal} {Phys. Rev. B}\
  }\textbf {\bibinfo {volume} {15}},\ \bibinfo {pages} {4476} (\bibinfo {year}
  {1977})}\BibitemShut {NoStop}%
\bibitem [{\citenamefont {Aharony}\ and\ \citenamefont
  {Imry}(1977)}]{aharonyimry1977}%
  \BibitemOpen
  \bibfield  {author} {\bibinfo {author} {\bibfnamefont {A.}~\bibnamefont
  {Aharony}}\ and\ \bibinfo {author} {\bibfnamefont {Y.}~\bibnamefont {Imry}},\
  }\href@noop {} {\bibfield  {journal} {\bibinfo  {journal} {J. Phys. C}\
  }\textbf {\bibinfo {volume} {10}},\ \bibinfo {pages} {L487} (\bibinfo {year}
  {1977})}\BibitemShut {NoStop}%
\bibitem [{\citenamefont {Schafer}\ and\ \citenamefont
  {Wegner}(1980)}]{wegnerschafer1980}%
  \BibitemOpen
  \bibfield  {author} {\bibinfo {author} {\bibfnamefont {L.}~\bibnamefont
  {Schafer}}\ and\ \bibinfo {author} {\bibfnamefont {F.}~\bibnamefont
  {Wegner}},\ }\href@noop {} {\bibfield  {journal} {\bibinfo  {journal} {Z.
  Phys. B}\ }\textbf {\bibinfo {volume} {38}},\ \bibinfo {pages} {113}
  (\bibinfo {year} {1980})}\BibitemShut {NoStop}%
\bibitem [{\citenamefont {Akhanjee}\ and\ \citenamefont
  {Rudnick}(2010)}]{spherePRL}%
  \BibitemOpen
  \bibfield  {author} {\bibinfo {author} {\bibfnamefont {S.}~\bibnamefont
  {Akhanjee}}\ and\ \bibinfo {author} {\bibfnamefont {J.}~\bibnamefont
  {Rudnick}},\ }\href@noop {} {\bibfield  {journal} {\bibinfo  {journal} {Phys.
  Rev. Lett.}\ }\textbf {\bibinfo {volume} {105}},\ \bibinfo {pages} {047206}
  (\bibinfo {year} {2010})}\BibitemShut {NoStop}%
\bibitem [{\citenamefont {Efetov}(1999)}]{efetovbook}%
  \BibitemOpen
  \bibfield  {author} {\bibinfo {author} {\bibfnamefont {K.}~\bibnamefont
  {Efetov}},\ }\href@noop {} {\emph {\bibinfo {title} {Supersymmetry in
  Disorder and Chaos}}}\ (\bibinfo  {publisher} {Cambridge University Press},\
  \bibinfo {address} {Cambridge},\ \bibinfo {year} {1999})\BibitemShut
  {NoStop}%
\bibitem [{\citenamefont {Kanzieper}(2002)}]{kanzieperPRL2002}%
  \BibitemOpen
  \bibfield  {author} {\bibinfo {author} {\bibfnamefont {E.}~\bibnamefont
  {Kanzieper}},\ }\href@noop {} {\bibfield  {journal} {\bibinfo  {journal}
  {Phys. Rev. Lett.}\ }\textbf {\bibinfo {volume} {89}},\ \bibinfo {pages}
  {250201} (\bibinfo {year} {2002})}\BibitemShut {NoStop}%
\bibitem [{\citenamefont {{Lerner}}(2003)}]{lerner}%
  \BibitemOpen
  \bibfield  {author} {\bibinfo {author} {\bibfnamefont {I.~V.}\ \bibnamefont
  {{Lerner}}},\ }\href@noop {} {\bibfield  {journal} {\bibinfo  {journal}
  {eprint arXiv:cond-mat/0307471}\ } (\bibinfo {year} {2003})}\BibitemShut
  {NoStop}%
\bibitem [{\citenamefont {Oppermann}\ and\ \citenamefont
  {Wegner}(1979)}]{oppermanwegner1979}%
  \BibitemOpen
  \bibfield  {author} {\bibinfo {author} {\bibfnamefont {R.}~\bibnamefont
  {Oppermann}}\ and\ \bibinfo {author} {\bibfnamefont {F.}~\bibnamefont
  {Wegner}},\ }\href@noop {} {\bibfield  {journal} {\bibinfo  {journal} {Z.
  Phys. B}\ }\textbf {\bibinfo {volume} {34}},\ \bibinfo {pages} {327}
  (\bibinfo {year} {1979})}\BibitemShut {NoStop}%
\bibitem [{\citenamefont {Edwards}\ and\ \citenamefont
  {Anderson}(1975)}]{edanderson}%
  \BibitemOpen
  \bibfield  {author} {\bibinfo {author} {\bibfnamefont {S.~F.}\ \bibnamefont
  {Edwards}}\ and\ \bibinfo {author} {\bibfnamefont {P.~W.}\ \bibnamefont
  {Anderson}},\ }\href@noop {} {\bibfield  {journal} {\bibinfo  {journal} {J.
  Phys. F}\ }\textbf {\bibinfo {volume} {5}},\ \bibinfo {pages} {965} (\bibinfo
  {year} {1975})}\BibitemShut {NoStop}%
\bibitem [{\citenamefont {Mehta}(2004)}]{mehta}%
  \BibitemOpen
  \bibfield  {author} {\bibinfo {author} {\bibfnamefont {M.~L.}\ \bibnamefont
  {Mehta}},\ }\href@noop {} {\emph {\bibinfo {title} {Random Matrices 3rd
  Edition}}}\ (\bibinfo  {publisher} {Elsevier},\ \bibinfo {address}
  {Amsterdam},\ \bibinfo {year} {2004})\BibitemShut {NoStop}%
\bibitem [{\citenamefont {Forrester}\ and\ \citenamefont
  {Witte}(2003)}]{FandW}%
  \BibitemOpen
  \bibfield  {author} {\bibinfo {author} {\bibfnamefont {P.~J.}\ \bibnamefont
  {Forrester}}\ and\ \bibinfo {author} {\bibfnamefont {N.~S.}\ \bibnamefont
  {Witte}},\ }\href@noop {} {\bibfield  {journal} {\bibinfo  {journal}
  {Nonlinearity}\ }\textbf {\bibinfo {volume} {16}},\ \bibinfo {pages} {1919}
  (\bibinfo {year} {2003})}\BibitemShut {NoStop}%
\bibitem [{\citenamefont {{Okamoto}}(1981)}]{okamoto1981}%
  \BibitemOpen
  \bibfield  {author} {\bibinfo {author} {\bibfnamefont {K.}~\bibnamefont
  {{Okamoto}}},\ }\href@noop {} {\bibfield  {journal} {\bibinfo  {journal}
  {Physica D}\ }\textbf {\bibinfo {volume} {2}},\ \bibinfo {pages} {525}
  (\bibinfo {year} {1981})}\BibitemShut {NoStop}%
\bibitem [{\citenamefont {{Kravtsov}}\ and\ \citenamefont
  {{Mirlin}}(1994)}]{kravtsov}%
  \BibitemOpen
  \bibfield  {author} {\bibinfo {author} {\bibfnamefont {V.~E.}\ \bibnamefont
  {{Kravtsov}}}\ and\ \bibinfo {author} {\bibfnamefont {A.~D.}\ \bibnamefont
  {{Mirlin}}},\ }\href@noop {} {\bibfield  {journal} {\bibinfo  {journal}
  {JETP}\ }\textbf {\bibinfo {volume} {60}},\ \bibinfo {pages} {656} (\bibinfo
  {year} {1994})}\BibitemShut {NoStop}%
\bibitem [{\citenamefont {Wegner}(1979{\natexlab{b}})}]{wegnerPRB1979}%
  \BibitemOpen
  \bibfield  {author} {\bibinfo {author} {\bibfnamefont {F.~J.}\ \bibnamefont
  {Wegner}},\ }\href@noop {} {\bibfield  {journal} {\bibinfo  {journal} {Phys.
  Rev. B}\ }\textbf {\bibinfo {volume} {19}},\ \bibinfo {pages} {783} (\bibinfo
  {year} {1979}{\natexlab{b}})}\BibitemShut {NoStop}%
\bibitem [{\citenamefont {de~Almeida}\ and\ \citenamefont
  {Thouless}(1978)}]{dealmeida}%
  \BibitemOpen
  \bibfield  {author} {\bibinfo {author} {\bibfnamefont {J.~R.~L.}\
  \bibnamefont {de~Almeida}}\ and\ \bibinfo {author} {\bibfnamefont {D.~J.}\
  \bibnamefont {Thouless}},\ }\href@noop {} {\bibfield  {journal} {\bibinfo
  {journal} {J. Phys. A}\ }\textbf {\bibinfo {volume} {11}},\ \bibinfo {pages}
  {983} (\bibinfo {year} {1978})}\BibitemShut {NoStop}%
\bibitem [{\citenamefont {Slevin}\ and\ \citenamefont
  {Ohtsuki}(1997)}]{slevinPRL1997}%
  \BibitemOpen
  \bibfield  {author} {\bibinfo {author} {\bibfnamefont {K.}~\bibnamefont
  {Slevin}}\ and\ \bibinfo {author} {\bibfnamefont {T.}~\bibnamefont
  {Ohtsuki}},\ }\href@noop {} {\bibfield  {journal} {\bibinfo  {journal} {Phys.
  Rev. Lett.}\ }\textbf {\bibinfo {volume} {78}},\ \bibinfo {pages} {4083}
  (\bibinfo {year} {1997})}\BibitemShut {NoStop}%
\bibitem [{\citenamefont {Parisi}(1979)}]{parisiPRL}%
  \BibitemOpen
  \bibfield  {author} {\bibinfo {author} {\bibfnamefont {G.}~\bibnamefont
  {Parisi}},\ }\href@noop {} {\bibfield  {journal} {\bibinfo  {journal} {Phys.
  Rev. Lett.}\ }\textbf {\bibinfo {volume} {43}},\ \bibinfo {pages} {1754}
  (\bibinfo {year} {1979})}\BibitemShut {NoStop}%
\bibitem [{\citenamefont {Hasan}\ and\ \citenamefont {Kane}(2010)}]{hasankane}%
  \BibitemOpen
  \bibfield  {author} {\bibinfo {author} {\bibfnamefont {M.~Z.}\ \bibnamefont
  {Hasan}}\ and\ \bibinfo {author} {\bibfnamefont {C.~L.}\ \bibnamefont
  {Kane}},\ }\href@noop {} {\bibfield  {journal} {\bibinfo  {journal} {Rev.
  Mod. Phys.}\ }\textbf {\bibinfo {volume} {82}},\ \bibinfo {pages} {3045}
  (\bibinfo {year} {2010})}\BibitemShut {NoStop}%
\bibitem [{\citenamefont {Pruisken}(1984)}]{pruiskenNUCB1984}%
  \BibitemOpen
  \bibfield  {author} {\bibinfo {author} {\bibfnamefont {A.~M.~M.}\
  \bibnamefont {Pruisken}},\ }\href@noop {} {\bibfield  {journal} {\bibinfo
  {journal} {Nucl. Phys. B}\ }\textbf {\bibinfo {volume} {235}},\ \bibinfo
  {pages} {277 } (\bibinfo {year} {1984})}\BibitemShut {NoStop}%
\end{thebibliography}
\end{document}